\documentclass[10pt, conference, compsocconf]{IEEEtran}
\usepackage{amsmath, amssymb, enumerate}
\usepackage{amsfonts}
\usepackage{amssymb}
\usepackage[usenames]{color}
\usepackage{graphicx}
\usepackage{times}
\usepackage{psfrag}
\usepackage{subfigure}
\usepackage{multirow}
\usepackage{epsfig}
\usepackage{epstopdf}
\usepackage[font=small,labelfont=bf]{caption}
\usepackage{url}
\usepackage{amsmath}

\makeatletter
\def\url@leostyle{%
  \@ifundefined{selectfont}{\def\UrlFont{\sf}}{\def\UrlFont{\small\ttfamily}}}
\makeatother

\begin{document}

\newcommand{\WCEI}{\mathit{WCEI}}

\title{Deterministic Real-time Thread Scheduling}
\author{ Heechul Yun, Cheolgi Kim and Lui~Sha\\
Department of Computer Science, \\
University of Illinois at Urbana-Champaign, Champaign, IL 61801\\
\{heechul,cheolgi,lrs\}@illinois.edu}
\maketitle
\thispagestyle{empty}

\begin{abstract}
Race condition is a timing sensitive problem. A significant source of timing
variation comes from non-deterministic hardware interactions such as cache misses.
While data race detectors and model checkers can check races, the enormous state
space of complex software makes it difficult to identify all of the races and those residual
implementation errors still remain a big challenge.
In this paper, we propose deterministic real-time scheduling methods to address scheduling
nondeterminism in uniprocessor systems.
The main idea is to use timing insensitive deterministic events, e.g, an instruction
counter, in conjunction with a real-time clock to schedule threads. By
introducing the concept of Worst Case Executable Instructions (WCEI),
we guarantee both determinism and real-time performance.
\end{abstract}

\section{Introduction}

% motivation
Software running on safety-critical embedded systems, such as avionics, medical
devices, and automotive systems requires high reliability because the consequences of
failure can be disastrous.  Even in non-safety-critical consumer electronics
such as smartphones, consumers demand higher reliability than ever as they play increasingly important roles in everyday life.

In multithreaded programs, thread interaction bugs, such as race condition, are sensitive to
interleaving patterns, so are hard to reproduce when thread scheduling is nondeterministic.
Such bugs can lead to the great challenge known as No Fault Found(NFF).
As it was observed that, ``Overall, better software has had a far greater
impact on reducing NFF than better hardware" by an avionics company \cite{NFF}
, software bugs have been observed to be increasingly major
causes of critical problems.% \cite{soderholm2007system}.
As noted by E. Lee \cite{lee2006problem}, correct reasoning of multithreaded programs is
extremely difficult because its output depends not only on the input but also on the thread schedules, which are essentially nondeterministic even in
uniprocessor systems.

As a solution for nondeterminism, a logical counter, such as an instruction counter, can be used to schedule the threads. If thread switching occurs on specific instruction counter values, the thread schedule is repeatable as long as the program has the same input.
Recent work in the parallel system community adapted this method to reduce nondeterminism in thread-scheduling of multi-core systems. These systems use a hardware instruction counter \cite{olszewski2009kendo}, or a compiler-generated virtual counter\cite{bergan2010coredet}, to control thread interleaving so that they produce deterministic schedules.

However, in real-time systems, we cannot solely rely on instruction counters,
because of the real-time constraints. In this paper, we
propose a novel thread-scheduling technique for uniprocessor systems that removes
time-dependant nondeterminism without sacrificing the real-time guarantee.
The key idea is that our scheduler uses both timer interrupts and instruction count
interrupts to schedule threads.
The traditional timer interrupt is used to keep up
with real-time. On the other hand, we also use an instruction counter to generate
an interrupt when a given number of instructions have been executed, to preserve
determinism in the scheduling decisions.

The challenge is to find a `good' deterministic counter and a mapping function between the counter value and real-time progress. If the mapping function is too pessimistic, the
task must be idle for a substantial amount of time, thus reducing CPU
utilization; if it is too aggressive, a high priority task can miss the
deadline because it has to wait a low priority task to execute all the assigned number of instructions.
This problem of finding a good mapping is challenging primarily because of the cache effect.
We evaluated our methods to find a good mapping function using a cycle-accurate processor simulator,
SimpleScalar. We also implemented a prototype RTOS with the proposed deterministic scheduler.

The rest of the paper is organized as follows. Section \ref{sec:motivation}
shows a motivating example. Section \ref{sec:drts} describes the deterministic
scheduler methodology. Section \ref{sec:impl} describes the prototype
implementation. Section \ref{sec:conclusion} concludes the paper.

\begin{figure*}[htp]
\begin{center}
  \includegraphics[width=12cm]{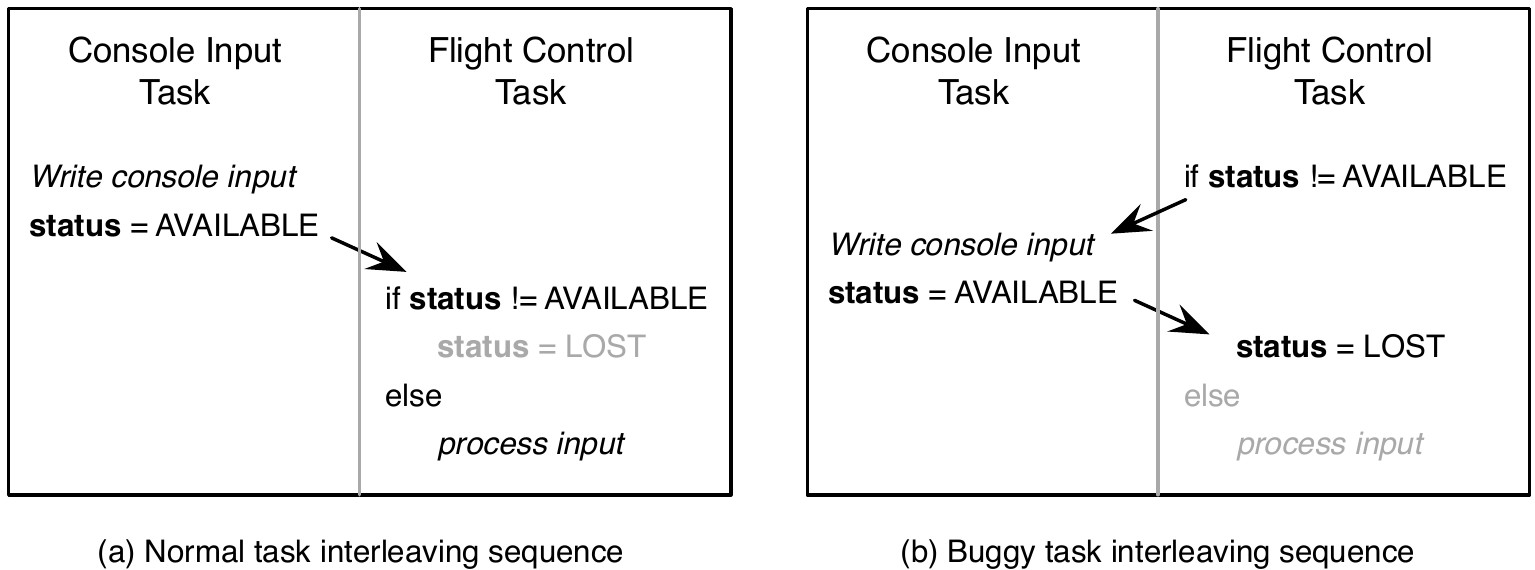}
  \caption{A race condition example simplified from paparazzi UAV source repository \cite{paparazzi}}\label{fig:buggy_read_diagram}.
\end{center}
\end{figure*}

\section{Motivating Example} \label{sec:motivation}

Race condition is a common mistake that is difficult to identify and fix. Consider, for example, Figure  \ref{fig:buggy_read_diagram}, which is found in an earlier version of the paparazzi \cite{paparazzi} UAV (Unmanned automatic vehicle) source code. Clearly, there is a race on reading and updating the \texttt{status} variable between the two threads. The left box is a correct run, while the right box is an incorrect one. Most of the time, this program performs fairly well. However, if the two threads interleave, as in the right box, the final \texttt{status} value is erroneously set to \texttt{LOST}.
While this bug can be removed easily by using a proper lock, finding the bug is not easy, because the bug is rarely manifested in practice. While there are many static and dynamic race detection tools \cite{pratikakis2006locksmith,savage1997eraser}, they
are often limited---they can miss or generate too many false positive bugs.
This kind of bug is a potential cause of NFF because its behavior is nondeterministic. By using the deterministic scheduler presented in this paper, however, it becomes deterministic and therefore easier to identify and fix.

\section{Deterministic Real-time Scheduling} \label{sec:drts}

\subsection{System Model and Definition }

We consider a real-time system that is modeled as a set of periodic real-time
tasks $\Gamma = \{ \tau_1, \tau_2, ..., \tau_n \}$ where $n$ is the number of
tasks.
The tasks can be \emph{dependent} on each other's operations and have a \emph{global shared memory} for inter-task communications.
The scheduling policy is rate monotonic: a higher priority task preempts a lower priority one when it arrives.
We assume there is a deterministic hardware counter (e.g., instruction counter) that is capable of measuring the program progress and generating an interrupt when it reaches a user-supplied number. We also assume that the deterministic counter and a system timer are the only sources of interruption in the system---threads are not allowed to use blocking system calls to wait for devices such as hard disks.

In a conventional fixed priority scheduler, when a task is preempted in the middle of execution by a higher priority task, the logical quantity of execution (the number of executed instructions) until the preemption point may vary because of the complex modern processor architecture, like cache and pipelining.
Such variation can cause races and NFF problems as presented in Section~\ref{sec:motivation}.

In our proposed deterministic scheduling, the scheduler uses a deterministic counter to ensure determinism in task executions. When a task $\tau$ is scheduled, the scheduler first foresees the time $t$ when $\tau$ must be scheduled out because of a higher priority task arrival or the completion of the task. The scheduler then estimates the number of instructions $I$ that can be executed for the duration $t$. After the task executes $I$ instructions, $\tau$ becomes idle, even when it finishes earlier than the estimated time $t$.
Notice that the task must not take longer than $t$ to execute $I$ instructions to avoid deadline miss. Hence, the next task can always be scheduled immediately on the arrival. Moreover, we assume that tasks execute for at least $T_{unit}$ once they are scheduled.

The key of deterministic scheduling is the estimation of the instruction quantity $I$ to be executed for the given duration $t$. We define the estimation function \emph{Worst-Case Executable Instructions (WCEI)}, because the execution of $I$ instructions must be guaranteed for the duration $t$. Notice that it must be worst-case estimation but also should not be too pessimistic for the sake of overall performance.
In Section \ref{sec:eval-detcnt}, we present how to estimate a good (tight) WCEI.

% The deterministic scheduling policy indeed schedules out a task early for determinism, which is called \emph{early yield}.
% Even if we have an ideal WCEI function, with which the task executes the estimated instructions exactly for the schedule duration in worst case, early yield can always happen if the environments, such as cache states, are friendly.
% However, an early yield from friendly environments does not degrade schedulability of the system.
% If WCET of a task is smaller than the total time budget for a period, deadline miss never occurs in a conventional scheduling policy.
% WCET is the execution time when all the parameters and the environments are the worst.
% If a task execution for a period is partitioned by high priority tasks, WCET will be hit only if all the scheduling fragments confront worst case scenario.
% An early yield because of friendly environments means that the system has successfully executed the worst-case amount of work given for the fragment.
% Consequently, such an early yield does not change the schedulability of the system.
% Thereby, the performance of deterministic scheduler fully depends on the WCEI estimation.

% TODO: early termination is ok. describe why

\subsection{An Example of Deterministic Scheduling}

\begin{figure}[htp]
\begin{center}
  \includegraphics[width=8cm]{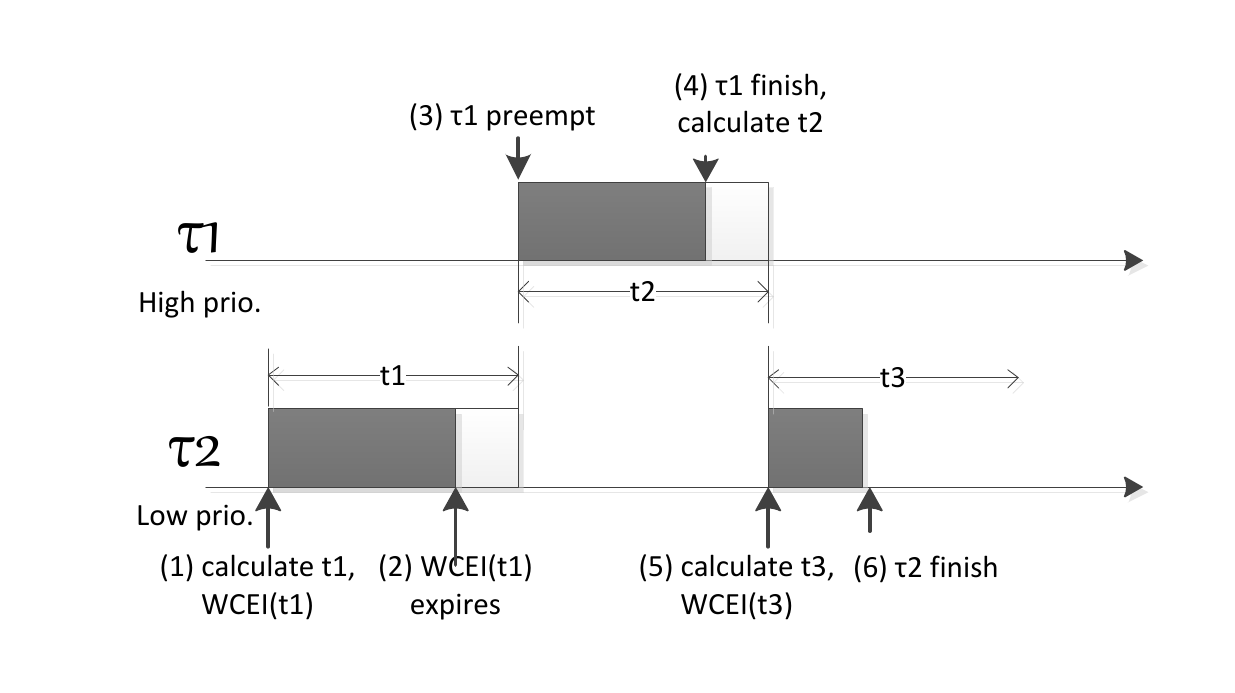}
  \caption{Task preemption in deterministic scheduling.} \label{fig:illustration}
\end{center}
\end{figure}

Figure \ref{fig:illustration} shows how the deterministic scheduling works. The system has two periodic tasks $\tau_1$ and $\tau_2$. (1) The scheduler schedules $\tau_1$ and computes $t_1$, the duration to the next scheduling event, and the corresponding WCEI. Then, it installs a interrupt which will be raised after executing WCEI instructions. (2) The interrupt is raised because the CPU executed all the instructions up to WCEI. Since the task finished earlier than $t_1$, it becomes idle. (3) $\tau_1$ arrives and is scheduled immediately. (4) When $\tau_2$ finish, the scheduler computes $t_2$ using a inverse of the WCEI function. (5) Similar to the step (1), the scheduler computes $t_3$ and the corresponding WCEI. (6) $\tau_2$ finishes before exhausting all WCEI instructions.

As this example clearly shows, WCEI computation plays an important role in the overall performance of deterministic scheduling; if the function is too pessimistic, the task must idle for a substantial amount of time hence reducing CPU utilization; if it is too aggressive, a higher priority task can be delayed by waiting for completion of the instruction budget of the lower priority task.
% In the next subsection, we will investigate how we can find a good WCEI.

\subsection{WCEI for Homogeneous Tasks}
\label{sec:eval-detcnt}

This section describes how to obtain a good WCEI function for homogeneous tasks in the form of
\begin{equation}
  \label{eq:wcei}
  \mathit{WCEI} = at - b
\end{equation}
where $t$ is the duration, $a$ is the execution speed, and $b$ is the cache cold-miss penalty.

The WCEI function correlates the executed instructions and the elapsed real-time. For an ideal processor, where every instruction takes exactly one cycle, the function would be simply the processor speed in Hz.
For any practical contemporary processor, however, the relationship is much more complicated for many reasons including cache effect and out-of-order execution. These are known to be difficult, if not impossible, to model.

We profiled the actual program execution using a cycle-accurate processor simulator, SimpleScalar, that modeled an alpha processor with cache and memory \cite{simplescalar} to obtain $a$ in Eq.~\eqref{eq:wcei}. Table \ref{tbl:simplescalar-params} shows the major system parameters we used in the simulator. Another possible way to profile such data is to use hardware performance counters found in most modern processors.

\begin{table}
\begin{center}
\caption{Simulator parameters. }
\begin{tabular}{|c|c|c|}
  \hline
  Module & Size & Latency  \\
  \hline
  L1-I\&D & 8KB & 2 cycles \\
  L2-uni  & 2MB & 16 cycles\\
  DRAM   & -  & 200,4 cycles\footnote{200 cycles for first chunk access, 4
cycles for inter-chunk access} \\
  \hline
\end{tabular}
\label{tbl:simplescalar-params}
\end{center}
\end{table}

\begin{figure}
 \includegraphics[width=8cm]{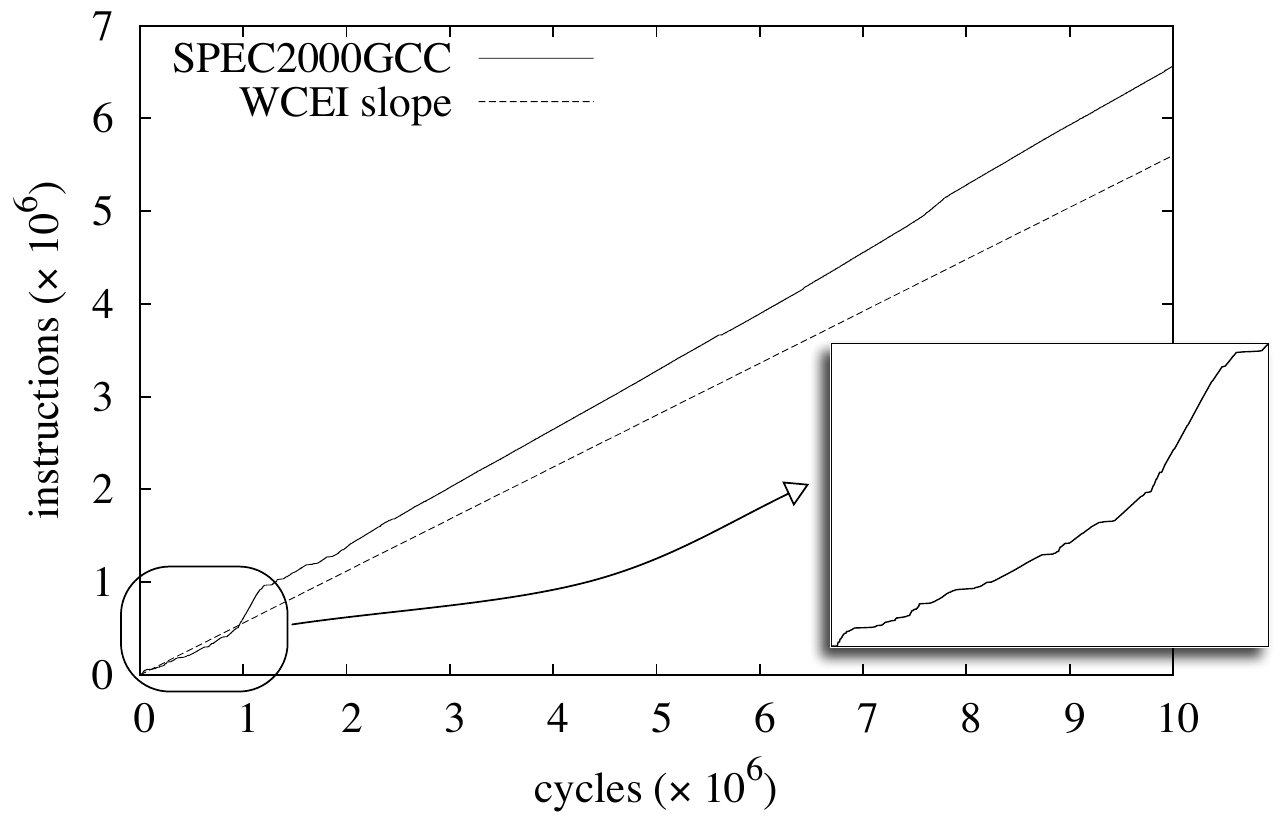}
  \caption{Execution profile and an WCEI example. WCEI is calculated assuming the $T_{unit}$ is 1M cycles (1ms in 1GHz CPU)}\label{fig:time_vs_inst}
\end{figure}

\begin{table}
\begin{center}
\caption{Utilization impact of unit time $T_{unit}$.}
\begin{tabular}{|c|c|c|c|}
  \hline
  $T_{unit}$ & WCEI rate & best rate & worst loss\\
  (M. cycles) & (inst/cycle) & (inst/cycle) & (\%) \\
  \hline
  1	&0.56	&0.89	&37.73 \\
  2	&0.59	&0.75	&21.52 \\
  3	&0.60	&0.71	&15.29 \\
  \hline
\end{tabular}
\label{tbl:unit_time}
\end{center}
\end{table}

We collected the processor cycle whenever an instruction was retired.
Figure \ref{fig:time_vs_inst} shows the retired instructions in terms of the elapsed cycles for a benchmark program, SPEC2000GCC, and an obtained WCEI function for this task.
The solid line shows the observed behavior, and the dotted line
presents coefficient $a$, which is obtained from the
smallest number of executed instructions over \emph{any} $T_{unit}$
, the smallest scheduling time unit during the program
execution. The choice of $T_{unit}$ is important: if it is too
short, the WCEI function becomes very conservative because the temporal
locality in the cache accesses varies significantly.
Table \ref{tbl:unit_time} shows the effect. As $T_{unit}$ increases
the worst case utilization loss is reduced. This is because the cache
locality is averaged over time. In all of our simulation, $T_{unit}$ was 1M cycles.
% Eq.~\eqref{eq:wcei} is completed by reflecting cache cold-miss penalty to $b$.

Note that if there are different program execution paths, we have to explore them all, and the WCEI must be the lower bound of them all. We argue that real-time control tasks often have very limited execution paths, therefore are systematically analyzable using automatic path exploration tools such as KLEE\cite{cadar2008klee}.

\subsection{WCEI for Multi-Phase Tasks}

Tasks are often divided into multiple phases of operation---for example, a computation phase and an update phase. In such cases, finding a single WCEI function, as described in the previous section, may result in a very pessimistic function because the instruction execution speed varies significantly over the different phases.

\begin{figure}
 \includegraphics[width=8cm]{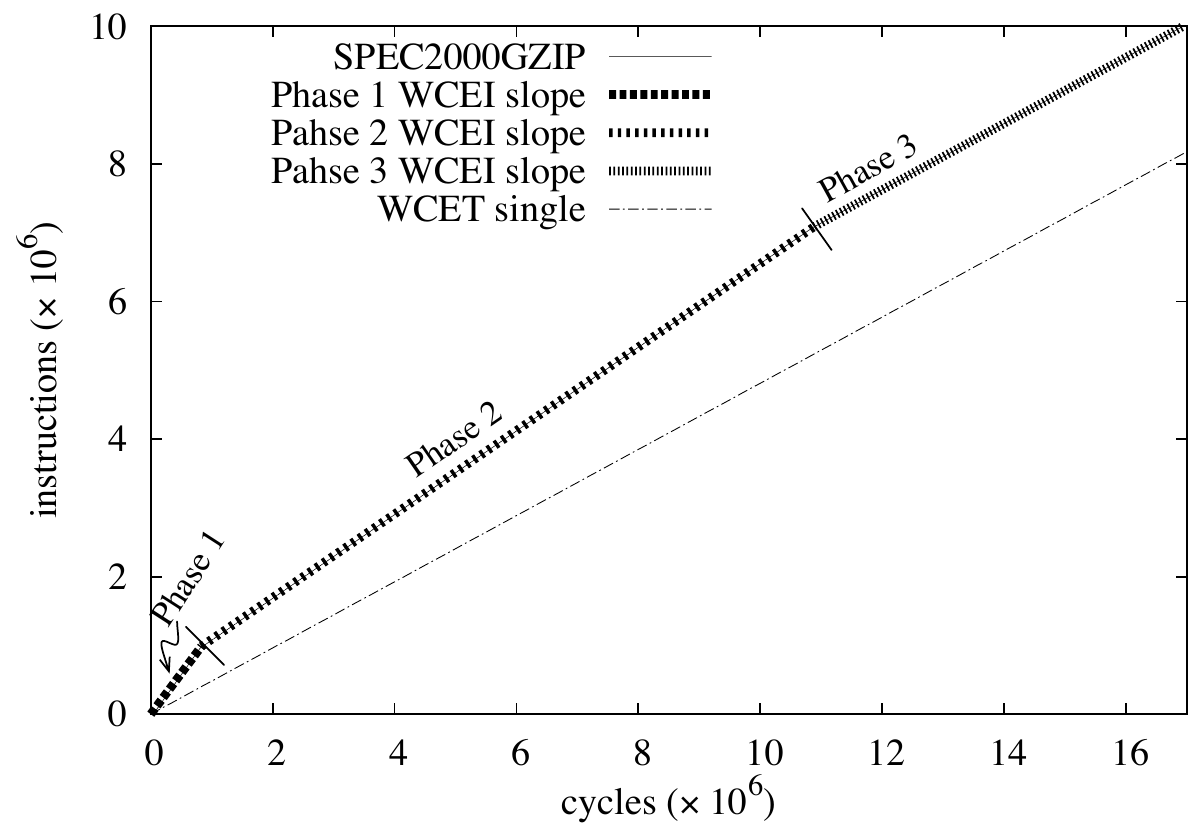}
 \caption{Execution profile for a multi-phase task.} \label{fig:multi}
\end{figure}

A solution to this problem is to identify a different WCEI function for each phase. In Figure \ref{fig:multi}, the profiled program SPEC2000GZIP shows three phases. Therefore, we computed three different WCEI functions: Phase 1, 2, and 3, which are tightly matched with the profiled execution behavior. In comparison, WCEI single, computed using the method described in the previous section, is not tight, i.e., the function wastes CPU. Table \ref{tbl:multi_uniform} compares the effectiveness of the multi-phase approach in terms of worst case CPU utilization losses. In the multi-phase approach, the utilization loss is very small---less than 3\% in all phases. In comparison, a single WCEI function results in up to 61.95\%, the worst case utilization loss.

Note that using the multi-phase function needs a means to notify the scheduler of the phase change so the scheduler can re-install the instruction counter interrupt using a new WCEI function.
This can be done manually, by inserting function calls in the application program, or automatically from the execution profile.
% similar to the method described
% \cite{pellizzoni2008coscheduling} to obtain a tight WCET bound over superblocks.

\begin{table}
\begin{center}
\caption{Utilization impact comparison of multi-phase WCEIs vs a single WCEI}
\begin{tabular}{|c|c|c|c|}
  \hline
  Region & WCEI rate & best rate & worst loss\\
        & (inst/cycle) & (inst/cycle) & (\%) \\
  \hline
  phase 1	&1.22	&1.25	&2.10\\
  phase 2	&0.61	&0.61	&0.28 \\
  phase 3   &0.48   &0.48   &0.61 \\
  \hline
  single	&0.48	&1.25	&61.95 \\
  \hline
\end{tabular}
\label{tbl:multi_uniform}
\end{center}
\end{table}

%\subsection{Mixed Real-time and Non Real-time Tasks}

\subsection{Mixed Real-time and Non Real-time Tasks}

While we described several optimization techniques in the previous sections, it is inevitable to lose some CPU utilization because the scheduler should idle after consuming the instruction budget defined by WCEI, which is supposed to be conservative. The idle cycles, however, can be utilized by other non-critical tasks, for which deadline misses are not an issue, as long as they are independent of the tasks under deterministic scheduling.
It is possible by selectively enabling the instruction counter based on the type of the task.

\section{Implementation} \label{sec:impl}

We implemented a preliminary prototype deterministic scheduler as a user level thread scheduler on top of Linux 2.6.36 running on an Intel Core2Duo processor based machine. Basic threading and synchronization APIs are the same as the standard pthread. We used two interrupts, timer and instruction counter, to implement the deterministic scheduler. For the instruction counter, we used the retired store instruction counter found in Intel Core2 Duo processor. We used perf\_events infrastructure \cite{perf_event} of Linux 2.6.36. Note that our current implementation only supports a single WCEI function.

% The code will be available at https://agora.cs.illinois.edu/display/realTimeSystems/Deterministic+Real-time+OS.
% For each periodic task arrival, it computes the time of next periodic task arrival event and its WCEI. Then it set the instruction counter generate an interrupt after the given instructions, WCEI, to be executed. In this way, preemption only occurs at the same location for a given execution path of the task.

\section{Conclusion and Future Work} \label{sec:conclusion}

In shared memory multi-thread systems, time-based preemptive scheduling is one of the main sources of nondeterminism. We proposed a counter-based deterministic scheduling method for periodic real-time tasks to eliminate such nondeterminism without violating the real-time property. A key to our approach was to design a good mapping function, called WCEI, that maps the counter and real-time.
Using a cycle-accurate processor simulator, we explored designs for WCEI functions and discussed related issues. We also made a prototype system showing that it is readily implementable in today's computer systems.

Future work will include evaluating a broad range of real-time applications using both a simulator and real hardware. Another interesting avenue would be to explore a deterministic hardware counter with different weights for different types of instructions (e.g., floating point, integer, and memory operations) or a scratchpad based MMU\cite{whitham2010studying}. Applying such hardware can potentially reduce pessimism in WCEI functions. 

\bibliographystyle{IEEEtran}
\bibliography{related,replay,dmt,racedetector}
\end{document}